\documentclass[prl,twocolumn,amsmath,amssymb,
superscriptaddress,floatfix]{revtex4-1}

\usepackage{graphicx}
\usepackage{color}
\usepackage{float}
\usepackage[normalem]{ulem}
\usepackage{hyperref}
\usepackage{bm}

\newcommand{\bk}{\mathbf{k}}
\newcommand{\bq}{\mathbf{q}}

\begin{document}
\title{Dynamical Planckian scaling of charge response at a particle-hole-asymmetric \\
quantum critical point with Kondo destruction}

\author{Ananth Kandala}
\thanks{These authors contributed equally to this work}
\affiliation{Department of Physics, University of Florida, Gainesville, Florida, 32611-8440, USA}
\author{Haoyu Hu}
\thanks{These authors contributed equally to this work}
\affiliation{Department of Physics \& Astronomy, Rice Center for Quantum Materials, Rice University, Houston, Texas 77005, USA}
\author{Qimiao Si}
\email{qmsi@rice.edu}
\affiliation{Department of Physics \& Astronomy, Rice Center for Quantum Materials, Rice University, Houston, Texas 77005, USA}
\author{Kevin Ingersent}
\email{ingersent@ufl.edu}
\affiliation{Department of Physics, University of Florida, Gainesville, Florida, 32611-8440, USA}

\begin{abstract}
Metallic quantum criticality is a central theme in a variety of strongly correlated systems.
Recent experiments have raised the fundamental question of how the charge response can be singular in cases where the Landau framework of quantum criticality allows singularity only in the spin channel.
Motivated by this emerging issue, we study the particle-hole-asymmetric regime of a Bose-Fermi Anderson model with power-law forms for both the bosonic bath spectrum and the fermionic band density of states. We realize a particle-hole-asymmetric quantum-critical state where quasiparticles are lost due to a critical destruction of Kondo screening, and demonstrate a dynamical Planckian scaling of the charge response.
Implications for a new regime of heavy-fermion quantum criticality and for Mott-Hubbard systems are discussed.
\end{abstract}

\maketitle

\textit{Introduction.}
Strange metals with a complete absence of quasiparticles are of extensive current interest in a variety of strongly correlated systems \cite{Kei17.1,Pas21.1,Phillips22}. Such states often arise near a quantum critical point (QCP) where one phase continuously transforms into another at absolute zero. The canonical case of antiferromagnetic (AF) metallic QCPs is traditionally formulated within the Landau framework of order-parameter fluctuations \cite{Hertz, Millis}. The majority of the Fermi surface does not experience the quantum-critical fluctuations at the ordering wave vector, so the associated quasiparticles retain their integrity. It is thus necessary to look beyond the Landau framework to understand strange metals. An alternative paradigm  incorporates additional quantum-critical degrees of freedom that destroy quasiparticles over the entire Fermi surface \cite{Si-Nature,Colemanetal,senthil2004a}. This happens when the so-called ``large" Fermi surface loses its quasiparticle weight and jumps to a ``small" Fermi surface as the system crosses the QCP; since the quantum phase transition is continuous, the quasiparticle weight vanishes everywhere on the Fermi surface. A variety of experiments, in a growing number of AF heavy-fermion metals \cite{paschen2004,Gegenwart2007,Friedemann.10,shishido2005,park-nature06,Kne08.1,Custers-2012,Mar19.1,Schroder,Kir20.2},
provide evidence for this class of Kondo-destruction QCP \cite{Si-Nature,Colemanetal,senthil2004a}.

A particularly intriguing recent development is the discovery of a singular critical charge response in YbRh$_2$Si$_2$ with dynamical Planckian ($\hbar\omega/k_\text{B} T$) scaling \cite{Prochaska20}. This behavior occurs near a QCP between AF and paramagnetic metallic phases, so it is incompatible with the Landau framework \cite{Hertz,Millis}, in which only the fluctuations of the order-parameter (here, the spin channel) should be singular.
Since YbRh$_2$Si$_2$ was one of the first heavy fermions to provide experimental evidence for a sudden jump of the Fermi surface across the QCP \cite{paschen2004,Gegenwart2007,Friedemann.10},
it is natural to associate the charge-channel singularity with the beyond-Landau nature of the quantum criticality. 
Such a charge-channel singularity is indeed found theoretically near Kondo-destruction transitions in Kondo-limit models with SU(2) symmetry \cite{Cai20} or very large spin degeneracy \cite{Zhu04,Komijani19}.
It has been argued that the singular charge response is important both for strange-metal behavior \cite{Prochaska20} and emergent high-$T_{\rm c}$ superconductivity \cite{Hu21.1},
making it important to establish the generality of the phenomenon. 
In the Kondo limit, which involves local moments
that are necessarily particle-hole-symmetric, the charge-channel singularity is ultimately traced to the fate of the local moments, and thus develops in the same energy range as the singular spin response \cite{Schroder,GrempelSi,ZhuGrempelSi,Glossop07,Zhu07}. What happens beyond away from this symmetric limit is an intriguing open question.

Recent experiments have pointed to a novel regime of quantum-critical heavy-fermion metals with inherent particle-hole asymmetry.
The quasi-kagome materials CeRhSn and CeIrSn show
strange-metal behavior \cite{Kim03,Tsuda18}, including a Gr\"{u}neisen ratio (of the thermal expansion to the specific heat) that diverges with an unusual exponent \cite{Tokiwa2015}, suggesting an entirely new universality class. 
Important clues about this quantum criticality are that CeRhSn is mixed valent \cite{Sundermann2021}, implicating the involvement of the charge degrees of freedom, but the differing responses to in-plane and out-of-plane pressure reveal local-moment degrees of freedom still to be well-defined \cite{Shimura2021}.
These systems motivate study of the interplay between intersite RKKY interactions and the local Kondo effect in entangled spin and charge channels.
Among the important open questions are whether there is a complete loss of quasiparticles, whether the charge and spin responses are both singular and, if so, whether the two singularities can acquire different energy scales, as indicated experimentally \cite{Shimura2021}.
Mixed valence is also expected to be important for other quantum-critical heavy-fermion metals such as 
$\beta$-YbAlB$_4$ \cite{Matsumoto_science11} and, likely, the $5f$-electron system PuCoGa$_5$ that superconducts with $T_c = 18.5$\,K (a record high for $f$-electron systems) \cite{Sarrao2002,Ramshaw2015.1}.
Theoretically, the role of valence fluctuations in heavy-fermion quantum criticality has been treated within the Landau framework \cite{Miyake2014}, but it has yet to be systematically studied in the context of beyond-Landau quantum criticality.

The relevant physics is captured by the Anderson lattice model in the particle-hole-asymmetric regime, incorporating RKKY interactions between the local spin degrees of freedom whose importance has been deduced from the aforementioned anisotropic pressure response of CeRhSn \cite{Shimura2021}.
A particularly powerful approach to elucidate the competing phases and quantum criticality of this model is extended dynamical mean-field theory \cite{Si.96,SmithSi-edmft,Chitra}, in which the Bose-Fermi Anderson (BFA) model appears as an effective Hamiltonian \footnote{See Supplemental Material for certain technical details and additional results.}.
In this work, we study BFA models with a sub-ohmic bosonic-bath spectrum and a fermionic band that has a power-law pseudogap in its density of states.
We realize an inherently particle-hole-asymmetric quantum-critical state where quasiparticles are lost due to a critical destruction of Kondo screening,
and demonstrate a dynamical Planckian ($\hbar \omega / k_\text{B} T$)
scaling of the charge response.

\textit{Model and solution methods.}
The BFA model is described by the Hamiltonian 
\begin{align}
\label{eq:Ham}
H_{\text{BFA}}
&= \epsilon_d \sum_\sigma n_{d\sigma} + U n_{d\uparrow} n_{d\downarrow} \notag \\
&\; + \sum_{\bk,\sigma} \epsilon_{\bk} c_{\bk\sigma}^{\dag} c_{\bk\sigma}
 + \frac{V}{\sqrt{N_{\bk}}} \sum_{\bk,\sigma} (d_\sigma^{\dag} c_{\bk\sigma} + \text{H.c.}) \\
&\; + \sum_{q,\mu} \omega_{\bq} \phi_{\bq\mu}^{\dag} \phi_{\bq\mu}
 + \sum_{\mu} \frac{g_{\mu} S_d^{\mu}}{\sqrt{N_{\bq}}}
   \sum_{\bq} (\phi_{\bq\mu}^{\dag} +\phi_{-\bq\mu}) , \notag
\end{align} 
where $d_{\sigma}^{\dag}$ creates an impurity electron with energy $\epsilon_d$ and
spin $z$ component $\sigma = \pm\frac{1}{2} \equiv \: \uparrow, \downarrow$;
$c_{k,\sigma}^\dag$ creates a conduction electron with wave vector $\bk$,
energy $\epsilon_{\bk}$, and spin $z$ component $\sigma$;
$\phi_{q\mu}^{\dag}$ creates a boson in bath $\mu \in \{x, y, z\}$ with wave vector $\bq$ and
energy $\omega_{\bq}$; and $S_d^{\mu} = \sum_{\sigma,\sigma'} d_{\sigma}^{\dag}
\frac{1}{2}\tau_{\sigma\sigma'}^{\mu} d_{\sigma'}$ is Cartesian component
$\mu$ of the impurity spin, $\tau^{\mu}$ being a Pauli matrix.
Other quantities entering Eq.\ \eqref{eq:Ham} are the Coulomb interaction $U$ between electrons
in the impurity level, the local hybridization $V$ between the impurity and conduction electrons,
and the coupling $g_{\mu}$ between component $\mu$ of the impurity's spin and the
corresponding bosonic bath. $N_{\bk}$ and $N_{\bq}$ are the number of $\bk$ and $\bq$ points,
respectively.
We take the electronic and bosonic densities of states to have the power-law forms  
\begin{align}
\label{rho_f}
\rho_f(\epsilon) &= \frac{1}{N_{\bk}} \sum_{\bk} \delta(\epsilon-\epsilon_{\bk})
  = \rho_{0,c} \Bigl|\frac{\epsilon}{D}\Bigr|^r c\biggl(\frac{|\epsilon|}{D}\biggr), \\
	\label{rho_b}
\rho_b(\omega) &= \frac{1}{N_{\bq}} \sum_q \delta(\omega-\omega_{\bq})
  = \rho_{0,b} \Bigl|\frac{\omega}{D}\Bigr|^s \Theta(\omega) \,
	  c\biggl(\frac{|\omega|}{D}\biggr),
\end{align}
where $D$ is an overall energy scale that we henceforth set to 1, $\rho_{0,b}$ and $\rho_{0,c}$ are normalization factors,
$\Theta(x)$ is the Heaviside function, and $c(x)$ is a cutoff function such that $c(x) \to 1$
for $x\ll 1$ and $c(x)\to 0$ for $x\to\infty$. The exponent $r > 0$ creates a power-law
pseudogap in the band density of states around the Fermi energy $\epsilon = 0$, while the
exponent $s<1$ describes a sub-ohmic dissipative bath.
The quantum-critical behavior of the purely-fermionic power-law Anderson model \cite{Withoff1990,Bulla1997,Gonzalez-Buxton1998,Logan2000} has been extensively investigated  \cite{Ingersent2002,Fritz2004,Pix12.1,Chowdhury2015},
but the previous study of its Bose-Fermi counterpart \cite{Pixley2013} primarily focused on the particle-hole-symmetric limit.
Here we present results obtained at particle-hole asymmetry using the recently-developed continuous-time quantum Monte Carlo (CTQMC) method for the spin-isotropic model \cite{Cai19.3} (which built on the general method \cite{Gull2011,Otsuki13})
as well as that for the Ising-anisotropic model \cite{pixley2011,Pixley2013}, and using the numerical renormalization-group (NRG) method \cite{Bulla2008} for the anisotropic case.

This paper focuses on the illustrative case of powers $r = 0.6$ and $s = 0.9$ with $\Gamma_0 = \pi\rho_{0,c}V^2=0.1$, $U = 0.03$, and $\epsilon_d\ne -U/2$ leading to breaking of particle-hole symmetry. We study (a) an Ising-anisotropic model with $g_\mu = g\delta_{\mu,z}$, describing bosons
coupling to just one component of the impurity spin, and (b) an SU(2)-symmetric model with
$g_{\mu} = g$ involving three equivalent bosonic baths $\mu = x$, $y$, $z$. Both variants of the
model have been solved using CTQMC with $\epsilon_d = -0.05$ and a cutoff function $g(x) = (2\pi)^{-1/2} \exp(-x^2/2)$.
We have also solved the Ising version using the NRG with $c(x) = \Theta(1-x)$ and discretization $\Lambda=9$; in these calculations, we generally chose $\epsilon_d\simeq -0.025$ to achieve entry into the quantum-critical regime at the lowest iteration number (or highest temperature) for the fixed $U$ and $\Gamma_0$ values.
The NRG and CTQMC are complementary in that the former method can reliably compute static
quantities arbitrarily close to absolute temperature $T=0$ and dynamics in the regime $|\omega|\gtrsim T$ \footnote{From this point on, we adopt units where $D = \hbar = k_\text{B} = g\mu_\text{B} = 1$.}, while the latter provides dynamical properties over all $\omega/T$ and is practical to apply in the SU(2)-symmetric case more relevant to many heavy-fermion materials.
We show that, despite their differing cutoffs, CTQMC and the NRG yield a consistent description of universal critical properties.

\begin{figure}[t!]
\centering
\includegraphics[width=\columnwidth]{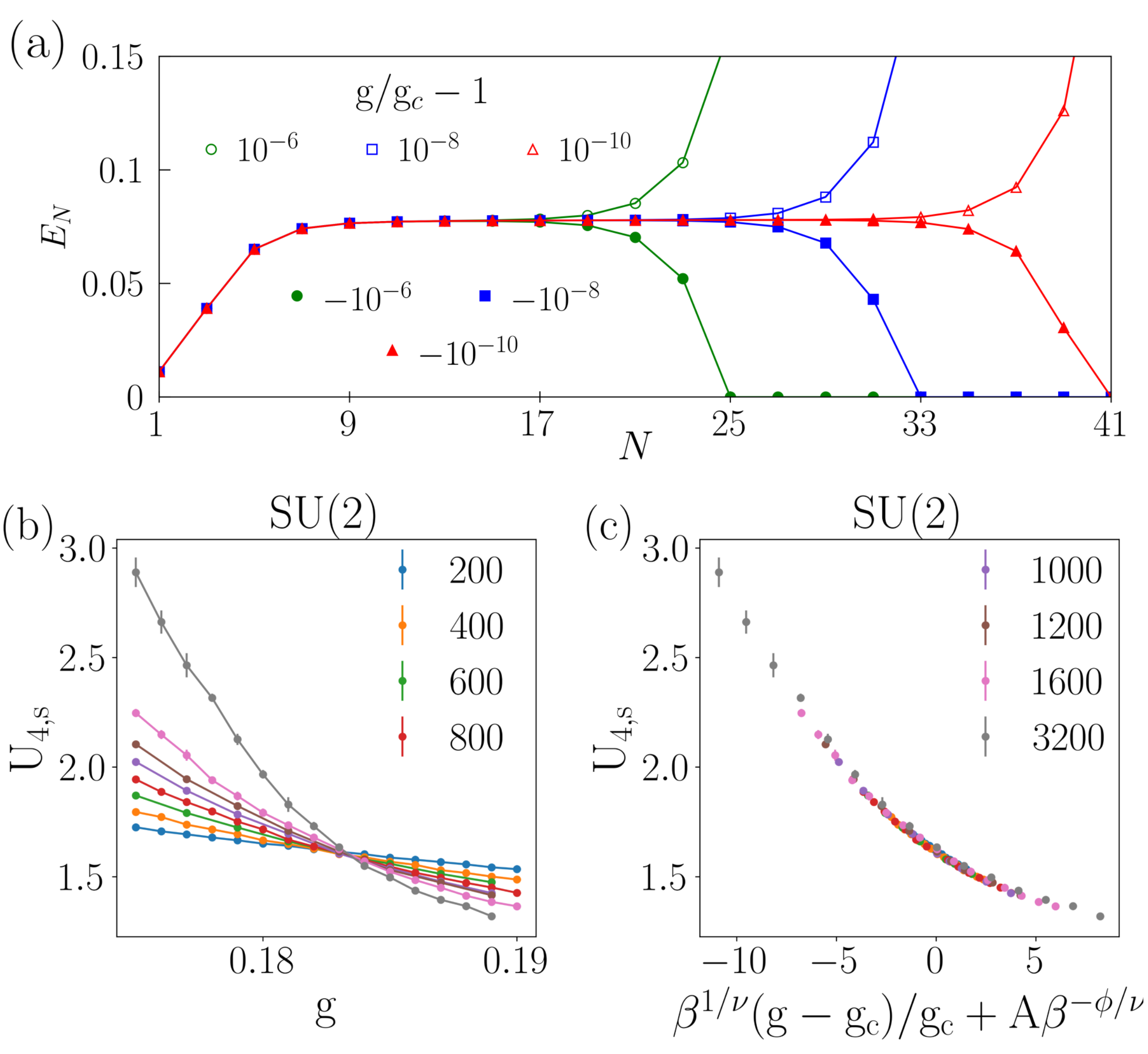}
\caption{\label{fig:find-QCP}%
(a) A low-lying NRG eigenvalue $E_N$ vs odd iteration number $N$ for bosonic couplings $g$ close to $g_c$ in the Ising BFA model. For $9\lesssim N\lesssim N^*(g)$, $E_N$ remains close to its value at the Kondo-destruction QCP. (b) Binder cumulant $U_{4,s}(\beta,g)$ vs $g$ for the SU(2) BFA model. (c) Optimized scaling collapse of the data from (b) using $g_c = 0.183(1), \nu^{-1}=0.684(15)$, $A=0.56$, and $\phi=0.50$. Legends are shared between (b), (c).}
\end{figure}

\textit{Loss of quasiparticles.}
When the Kondo effect occurs, the low-energy physics is characterized by quasiparticles with small but nonzero weight. The quasiparticle weight $Z$ is proportional to the energy scale associated with the development of the composite fermions.
 We start by searching for a Kondo-destruction QCP in the Ising-anisotropic BFA model using the NRG method.
The NRG provides a discrete approximation to the many-body spectrum at a sequence of energy scales $D_N \simeq D\Lambda^{-N/2}$, describing the
essential physics at temperatures $T \simeq D_N$ for $N = 0$, $1$, $2$, $\ldots$.
We locate the Kondo-destruction QCP by determining for trial values of $g$ whether the spectrum approaches, for iterations $N\gtrsim N^*(g)$, that of the particle-hole-asymmetric Kondo fixed point
(for $g < g_c$) or the bosonic-localized (Kondo-destroyed) fixed point (for $g > g_c$); see Fig.\ \ref{fig:find-QCP}(a).
A process of bisection on the $g$ axis increases $N^*$ and allows $g_c$ to be determined to one part in $10^{12}$ or better. At $g_c$, the energy scale for the development of the well-defined Kondo singlet has gone to zero. This means that the quasiparticle weight $Z$ vanishes. 
The loss of quasiparticles is also captured in the NRG many-body spectrum at $g_c$ \cite{Note1}, which no longer has a Fermi-liquid description.

The Kondo-destruction QCP can be located within the CTQMC approach by constructing the Binder cumulant
$U_{4,s} = \langle M_z^4\rangle / \langle M_z^2\rangle$, where
$\langle M_z^n\rangle = \langle [\beta^{-1} \int_0^{\beta} S_d^z(\tau) d\tau]^n\rangle$ with
$\beta=1/k_B T$.
Near a QCP at $g=g_c$, one expects $U_{4,s}(\beta,g) = f(\beta^{1/\nu}(g-g_c)/g_c + A\beta^{-\phi/\nu})$, where $f(x)$ is a scaling function and $A\beta^{-\phi/\nu}$ represents the subleading correction to scaling.
The Ising-anisotropic model is addressed in Fig.\ S1, while Fig.\ \ref{fig:find-QCP}(b) shows raw data for the SU(2)-symmetric model. 
One can identify $g_c$ as the point where $U_{4,s}$ vs $g$ curves for different values of $\beta$ cross at a single point.
Figure \ref{fig:find-QCP}(c) presents the result of optimizing the scaling collapse of the data from Fig.\ \ref{fig:find-QCP}(b) with respect to trial values of $g_c$, $\nu$, $A$, and $\phi$. As we will show in the next section, there is no Pauli form at $g_c$ for either the spin or charge susceptibility even at vanishingly small temperatures. This provides an alternative way to characterize the absence of Landau quasiparticles at the Kondo-destruction QCP.

\begin{figure}[t!]
\centering
\includegraphics[width=\columnwidth]{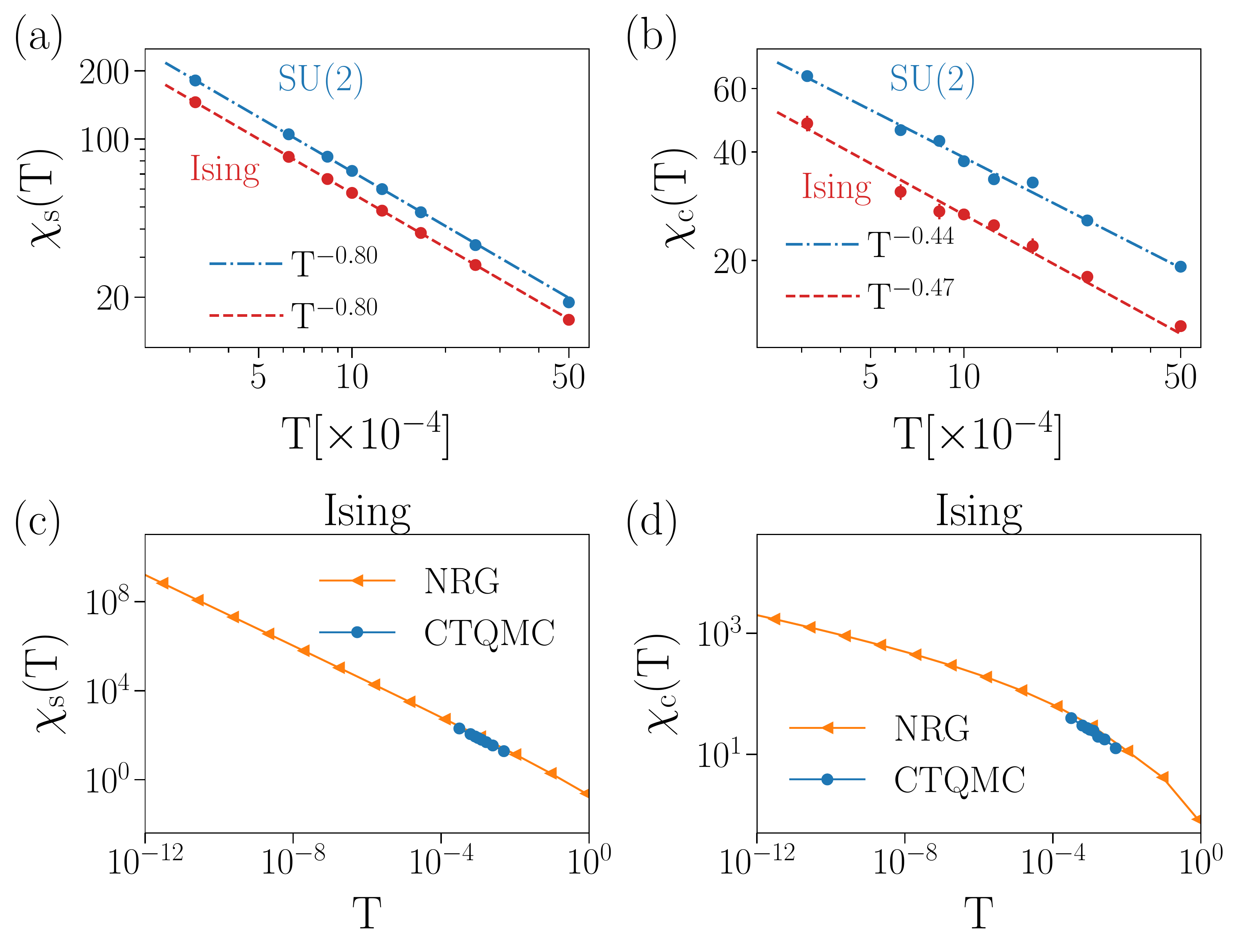}
\caption{\label{fig:chi_static}
Temperature dependence of static local susceptibilities at the Kondo-destruction QCP: (a) spin and (b) charge
susceptibilities for the Ising and SU(2) BFA models, computed using CTQMC. The SU(2) charge
data are multiplied by $1.5$ for clarity. 
(c) and (d) respectively superimpose Ising CTQMC spin and charge results on their NRG counterparts. All data for $\epsilon_d = -0.05$.}
\end{figure}

\textit{Singular charge and spin responses.}
To probe the critical properties, we first examine the static local spin and charge susceptibilities
\begin{align} 
\chi_s(T)& = -\biggl. \frac{\partial\langle S_d^z\rangle}{\partial h}\biggr|_{h=0}
           = \int_0^\beta \langle S_d^z(\tau)S_d^z(0) \rangle \, d\tau, \\
\chi_c(T)& = -\frac{\partial\langle n_d\rangle}{\partial\epsilon_d}
           = \int_0^\beta \langle :\!n_d(\tau)\!: \; :\!n_d(0)\!:\rangle \, d\tau, 
\end{align}
where $:\!n_d\!: \: = \sum_\sigma n_{d,\sigma} - \langle \sum_\sigma n_{d,\sigma}\rangle$
and $h$ is a local magnetic field that enters the Hamiltonian through a term $h S_d^z$.

As shown in Fig.\ \ref{fig:chi_static}, at the QCP both $\chi_s$ and $\chi_c$ have divergent
temperature dependences:
\begin{equation}
\label{eq:chi_static}
\chi_{\alpha}(T) = C_{\alpha} \, T^{-x_{\alpha}} \quad \text{for } \alpha = s,\, c .
\end{equation}
CTQMC gives $x_s=0.80(1)$ for both the Ising and SU(2) BFA models [Fig.\ \ref{fig:chi_static}(a)],
consistent within the numerical uncertainty with the NRG result $x_s=0.7906(1)$ for the Ising BFA model.
Figure \ref{fig:chi_static}(c), plotted over a much wider temperature range, shows that CTQMC
closely reproduces an NRG $\chi_s(T)$ (calculated for the same values of $U$, $\epsilon_d$, and $\Gamma_0$) that is already approaching its asymptotic low-temperature
behavior for $T \lesssim 0.01D$.

In the charge sector, CTQMC gives $x_c = 0.44(3)$ for the Ising BFA model and $x_c = 0.47(2)$ for
its SU(2) counterpart [Fig.\ \eqref{fig:chi_static}(b)], whereas the NRG Ising BFA result is
$x_c=0.1164(1)$. Figure \ref{fig:chi_static}(d) reveals that the CTQMC and NRG data
coincide closely, but $\chi_c(T)$ reaches its asymptotic regime only at
temperatures below those accessible to CTQMC. This behavior can be understood by considering
corrections to Eq.\ \eqref{eq:chi_static}, i.e.,
$1/\chi_{\alpha}(T) = C_{\alpha}^{-1} \, T^{x_{\alpha}} +
(C'_{\alpha})^{-1} \, T^{x'_\alpha} + \ldots$,
where $x'_\alpha > x_\alpha$.
The leading correction is smaller than $\epsilon$ times the asymptotic term in Eq.\ \eqref{eq:chi_static} for temperatures
$T < |\epsilon C'_\alpha/C_\alpha|^{1/[x'_\alpha-x_\alpha]}$. The NRG data yield
$x'_s = 0.90(4)$, $C'_s/C_s \approx 30$, $x'_c = 0.24(1)$, and $C'_c/C_c \approx 0.4$.
The facts that $x'_c - x_c \ll x'_s - x_s$ and $C'_c/C_c \ll C'_s/C_s$ make it necessary to go to much lower temperatures to access the asymptotic behavior of $\chi_c$.

\begin{figure}[t!]
\centering\mbox{\includegraphics[width=0.98\columnwidth]{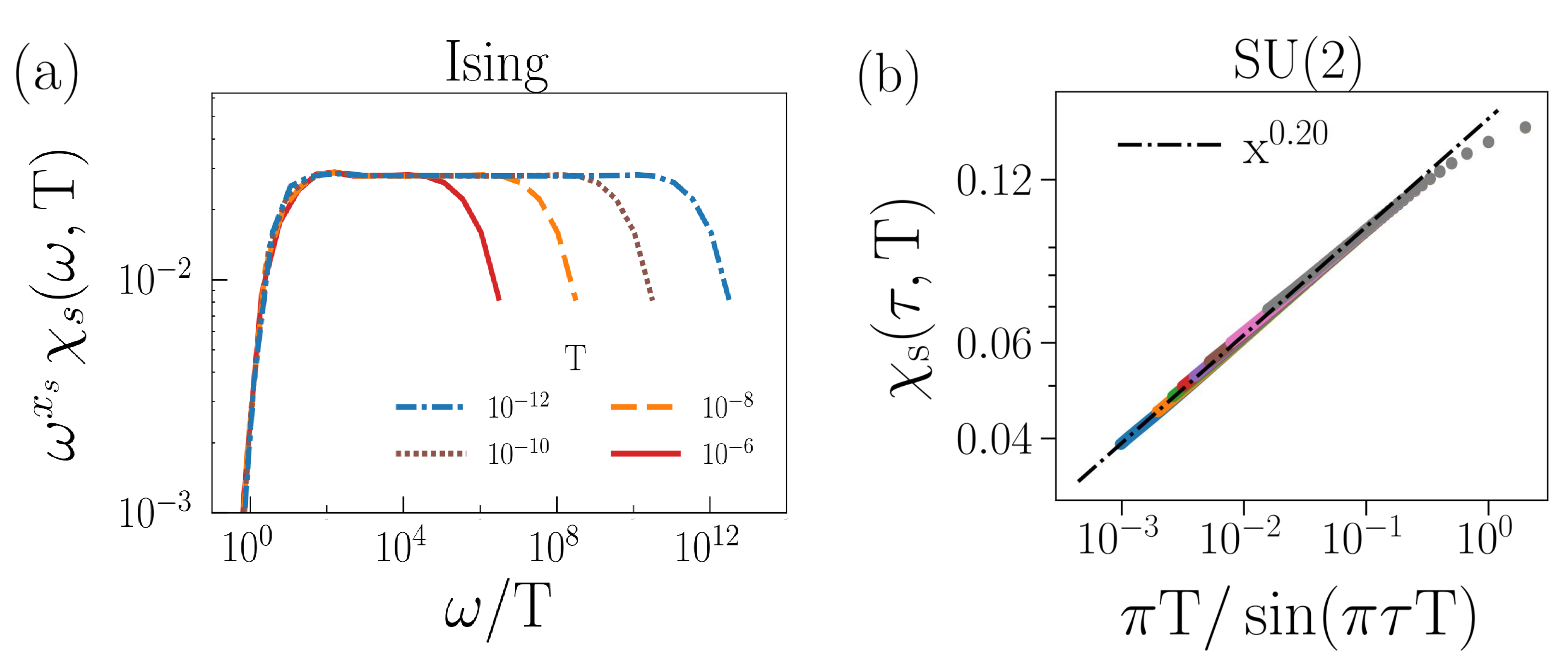}}
\caption{\label{fig:chi_dynamic_spin}%
Dynamical local spin susceptibility $\chi_s$ at the Kondo-destruction QCP:
(a) Real-frequency Ising NRG data at different temperatures $T$ collapse when plotted vs $\omega/T$.
(b) Imaginary-time SU(2) CTQMC data at different inverse temperatures $\beta$ collapse with dynamical exponent $0.2000(1)$ when plotted vs $\pi T/\sin(\pi \tau T)$. Colors in (b) are as specified in Figs.\ \ref{fig:find-QCP}(b) and \ref{fig:find-QCP}(c).
}
\end{figure}

\textit{Dynamical Planckian scaling of critical spin and charge responses.}
We now consider the dynamical responses that accompany the loss of quasiparticles at the Kondo-destruction QCP, beginning with the dynamical spin
susceptibility, which in imaginary time is
$\chi_{s}(\tau,T) = \langle S^z(\tau) S^z(0)\rangle - \langle S^z(0)\rangle^2$.
Figure \ref{fig:chi_dynamic_spin}(a) of the Ising case shows that real-frequency NRG data over 6 decades of temperature and more than 8 decades of frequency collapse onto a single curve, demonstrating a Planckian ($\omega/T$) scaling \footnote{The NRG does not capture the correct scaling form of the scaling function for $\omega\ll T$ due to truncation errors; see, e.g., Ref. \onlinecite{Ingersent2002}.}. For $\omega\gg T$, $\chi_s(\omega,T) \sim \omega^{-x_s}$ up to $\omega\simeq 0.01D$, where deviations from scaling set in.
Dynamical scaling also occurs in the SU(2) case, as demonstrated by the imaginary-time dependence of the spin susceptibility from CTQMC.
The power-law dependence on
$\pi T/\sin(\pi \tau T)$, shown in Fig.\ \ref{fig:chi_dynamic_spin}(b),
describes a scaling collapse of the susceptibility in terms of $\omega_n/T$ \cite{Note1}.

\begin{figure}[t!]
\centering\mbox{\includegraphics[width=0.98\columnwidth]{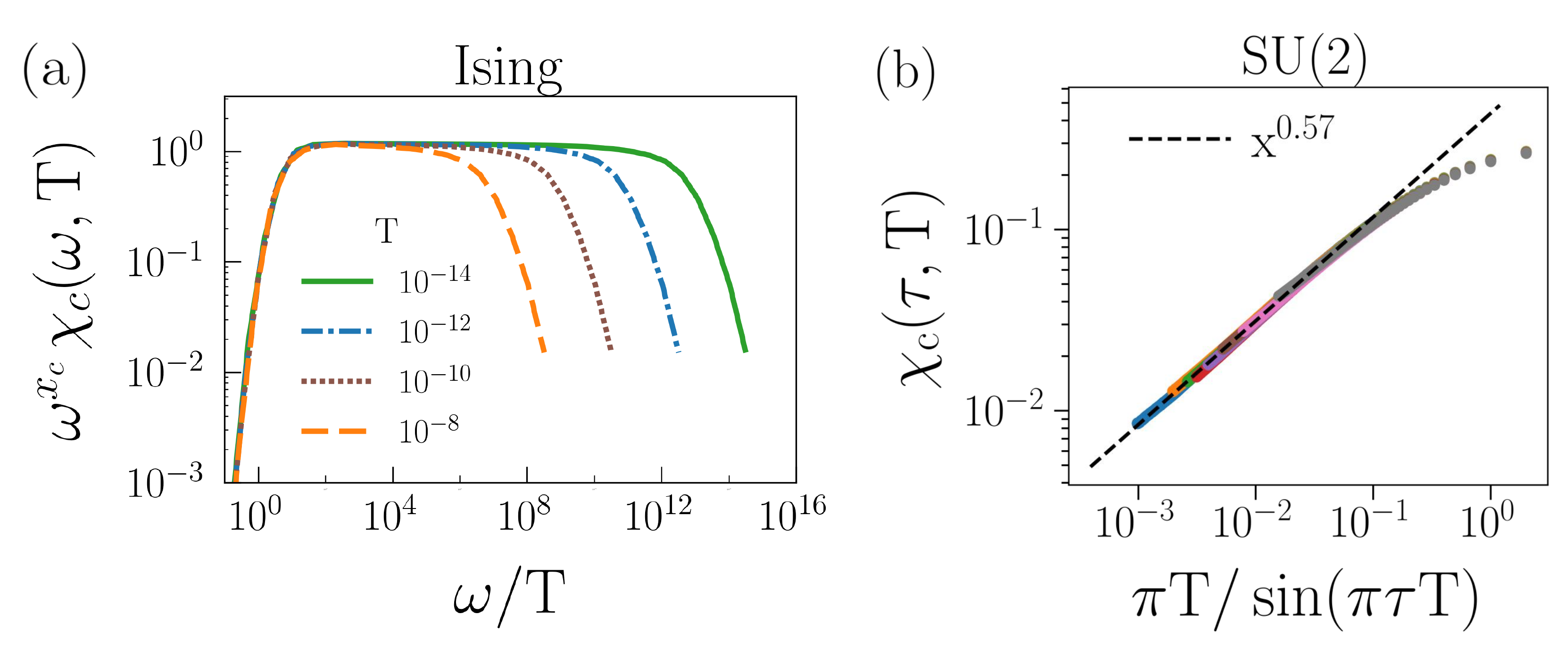}}    
\caption{\label{fig:chi_dynamic_charge}%
Dynamical local charge susceptibility $\chi_c$ at the Kondo-destruction QCP:
(a) Real-frequency Ising NRG data at different temperatures $T$ collapse when plotted vs $\omega/T$.
(b) Imaginary-time SU(2) CTQMC data at different inverse temperatures $\beta$ collapse with dynamical exponent $0.5728(2)$ when plotted vs $\pi T/\sin(\pi \tau T)$. Colors in (b) are as specified in Figs.\ \ref{fig:find-QCP}(b) and \ref{fig:find-QCP}(c).
}
\end{figure}

Finally, we turn to the principal focus of this work, namely the dynamical critical charge response probed by $\chi_c(\tau,T) = \langle :\!n_{d}(\tau)\!: \; :\!n_{d}(0)\!:\rangle$.
Figure \ref{fig:chi_dynamic_charge}(a) 
shows the real-frequency dynamics in the Ising case. NRG data spanning 6 decades of temperature and more than 10 decades of frequency show a clear 
scaling collapse in terms of $\omega/T$. For $\omega\gg T$, $\chi_c(\omega,T) \sim \omega^{-x_c}$, but corrections to scaling set in at lower frequencies than for $\chi_s(\omega,T)$, consistent with the greater prominence of subleading terms in $\chi_c(T)$ [see Fig.\ \ref{fig:chi_static}(d)].
This result is also corroborated by the CTQMC result for the SU(2) case, as shown in Fig.\,\ref{fig:chi_dynamic_charge}(b). The imaginary-time dependence corresponds to an $\hbar \omega_n /k_B T$ scaling of the dynamical charge susceptibility.
Moreover, the cutoff scale for the scaling region is considerably different between the charge and spin susceptibilities, demonstrating that the particle-hole asymmetry has liberated the scaling of the charge response from its spin counterpart.

\textit{Discussion.}
Several remarks are in order. 
First, our results show remarkable consistency between the quantum-critical behavior in the particle-hole-asymmetric
Bose-Fermi Anderson model in the Ising and SU(2) cases. This is to be contrasted with the particle-hole-symmetric case,
where a recent study shows that the SU(2) model can have
important differences from its Ising counterpart \cite{Cai19.3}. The consistency between 
the two spin symmetries allows us to reach a very complete understanding of the quantum-critical behavior, 
demonstrating the loss of quasiparticles and uncovering dynamical Planckian scaling. Our results motivate further studies for other combinations of the key parameters of the model: the pseudogap power $r$ of the fermionic band and the power-law exponent $s$ of the bosonic bath.

Second, the particle-hole-asymmetric Bose-Fermi Anderson model represents, through the EDMFT framework, an effective description of the 
Anderson lattice model in the mixed-valent regime. Our finding of the dynamical Planckian 
scaling of the charge responses at the Kondo-destruction QCP implicates a singular charge response in a new class of heavy-fermion QCPs. Optical conductivity measurements in the quantum-critical regime of CeRhIn and CeIrIn offer an important avenue for progress. Another way to elucidate this new regime of heavy-fermion quantum criticality will be to ascertain the degree to which the temperature and energy windows for dynamical scaling are different between the charge and spin channels.

In the heavy-fermion compound
$\beta$-YbBAl$_4$, evidence has been advanced for a critical charge mode in the quantum-critical regime \cite{Kobayashi2022}. Our work motivates the examination of the associated charge responses as a function of both frequency and temperature.

Third, the Bose-Fermi Anderson model also characterizes the one-band Hubbard model with explicit intersite exchange interactions, i.e., the $t$-$J$-$U$ Hubbard model \cite{SmithSi-edmft}. Away from half-filling, the model is particle-hole-asymmetric. 
A similar effective model also arises when the $J$ interactions are randomly distributed \cite{Chowdhury2021,Joshi2020}. In this way,
our determination of the dynamical Planckian scaling of the charge response sheds light on both the singular density
fluctuations, as observed in the electron energy-loss spectrum (EELS) in Bi-2212 \cite{Mitrano18},
and the related singularities as revealed 
by the $\hbar \omega / k_\text{B} T$ scaling of the optical conductivity \cite{Mar03.1}.

\textit{Summary.} 
We have shown that the Bose-Fermi Anderson model serves as an exemplary setting for particle-hole-asymmetric quantum criticality, in which the quasiparticles are lost due to critical Kondo destruction.
We have uncovered a dynamical Planckian scaling arising from the quasiparticle loss.
Our results deepen understanding of the emerging physics of singular charge response in beyond-Landau metallic quantum
critical points, and identify a new regime of heavy-fermion quantum criticality in which to explore this physics.
Furthermore, our work provides new insights into the emerging charge-channel singularities near the optimal carrier 
doping of Mott-Hubbard systems.

We thank Silke Paschen and D.\ T.\ Adroja for useful discussions. Work at Rice University was primarily supported by the NSF Grant No. DMR-1920740 and by the Robert A.\ Welch Foundation Grant No. C-1411 (H.H. and Q.S.). 
Computational work has used the 
Shared University Grid at Rice funded by NSF under Grant EIA-0216467, a partnership between
Rice University, Sun Microsystems, and Sigma Solutions, Inc., the Big-Data Private-Cloud 
Research Cyberinfrastructure MRI-award funded by NSF under Grant No. CNS-1338099 and by
Rice University, and the Extreme Science and Engineering Discovery Environment (XSEDE) by NSF
under Grant No. DMR170109. One of us (Q.S.) acknowledges the hospitality of the Aspen Center for
Physics, which is supported by NSF grant No. PHY-1607611.

\bibliography{pha-bfam}
\end{document}


\section*{Supplemental Material}
\vspace*{-2ex}
\begin{center}
\textbf{
Dynamical Planckian scaling of charge response at a particle-hole-asymmetric \\
quantum critical point with Kondo destruction} \\[2ex]

Ananth Kandala, Haoyu Hu, Qimiao Si, and Kevin Ingersent
\end{center}

\subsection{Background: Bose-Fermi Anderson Model via Extended Dynamical Mean-Field Theory}

The Bose-Fermi Anderson (BFA) impurity model has been widely used as an effective model to study the heavy-fermion systems \cite{Si-JPSJ2014}. It can be derived from the conventional Anderson lattice model via extended dynamical mean-field theory (EDMFT) \cite{Si.96,SmithSi-edmft,Chitra}. A generic Anderson lattice model with RKKY interactions can be described by the following Hamiltonian 
\begin{eqnarray}
H &=& \sum_{ij,\sigma}t_{ij}c_{i,\sigma}^\dag c_{j,\sigma} + \sum_i V(c_{i,\sigma}^\dag d_{i,\sigma} +\text{h.c.}) +\sum_{ij}I_{ij}\bm{S}_i\cdot \bm{S}_j 
+ \epsilon_d \sum_{i,\sigma}d_{i,\sigma}^\dag d_{i,\sigma} +
U\sum_{i}n_{i,\uparrow}n_{i,\downarrow}
\end{eqnarray} 
where $c^\dag_{i,\sigma}$ ($d^\dag_{i,\sigma}$) creates a conduction (localized) electron with spin $\sigma$ at lattice site $i$, while $t_{ij}$ and $I_{ij}$ are the conduction-electron hopping and the RKKY local-moment exchange interaction, respectively; $V$ is the on-site hybridization between conduction and localized electrons, $U$ is the Coulomb interaction between localized electrons on the same site, and $\epsilon_d$ is the level energy of $d$ electrons; $\bm{S}_i = \sum_{\sigma,\sigma'} d_{i,\sigma}^\dag \frac{1}{2} \bm{\sigma}d_{i,\sigma'}$ and ${n}_i = \sum_\sigma d_{i,\sigma}^\dag d_{i,\sigma}$ are the $d$-electron spin operator and number operator respectively. EDMFT maps the lattice model to a Bose-Fermi Anderson single-impurity model [Eq.\ (1) of the main text] describing a representative lattice site where $d$ electrons hybridize with a fermionic band representing the effect of the conduction band and are coupled via their spin $\bm{S}_d$ to a dissipative bosonic bath representing spin fluctuations caused by RKKY interactions with local moments at other lattice sites.

\subsection{NRG Critical Spectrum}

NRG results presented in the main paper and below were computed for discretization $\Lambda = 9$ using a basis of up to 8 bosons per site of the bosonic Wilson chain, and retaining up to 1000 many-body eigenstates after each iteration. The hybridization width used in the calculations was $\Gamma_\text{NRG} = A(\Lambda,r) \Gamma$, where $\Gamma = 0.1 D$ and $A(\Lambda,r)$ is a multiplicative factor---such that $A(9,0.6)\simeq 1.739$---introduced to compensate for band discretization effects \cite{Gonzalez-Buxton1998}.

\begin{figure}[h!]
\centering\mbox{\includegraphics[width=0.45\columnwidth]{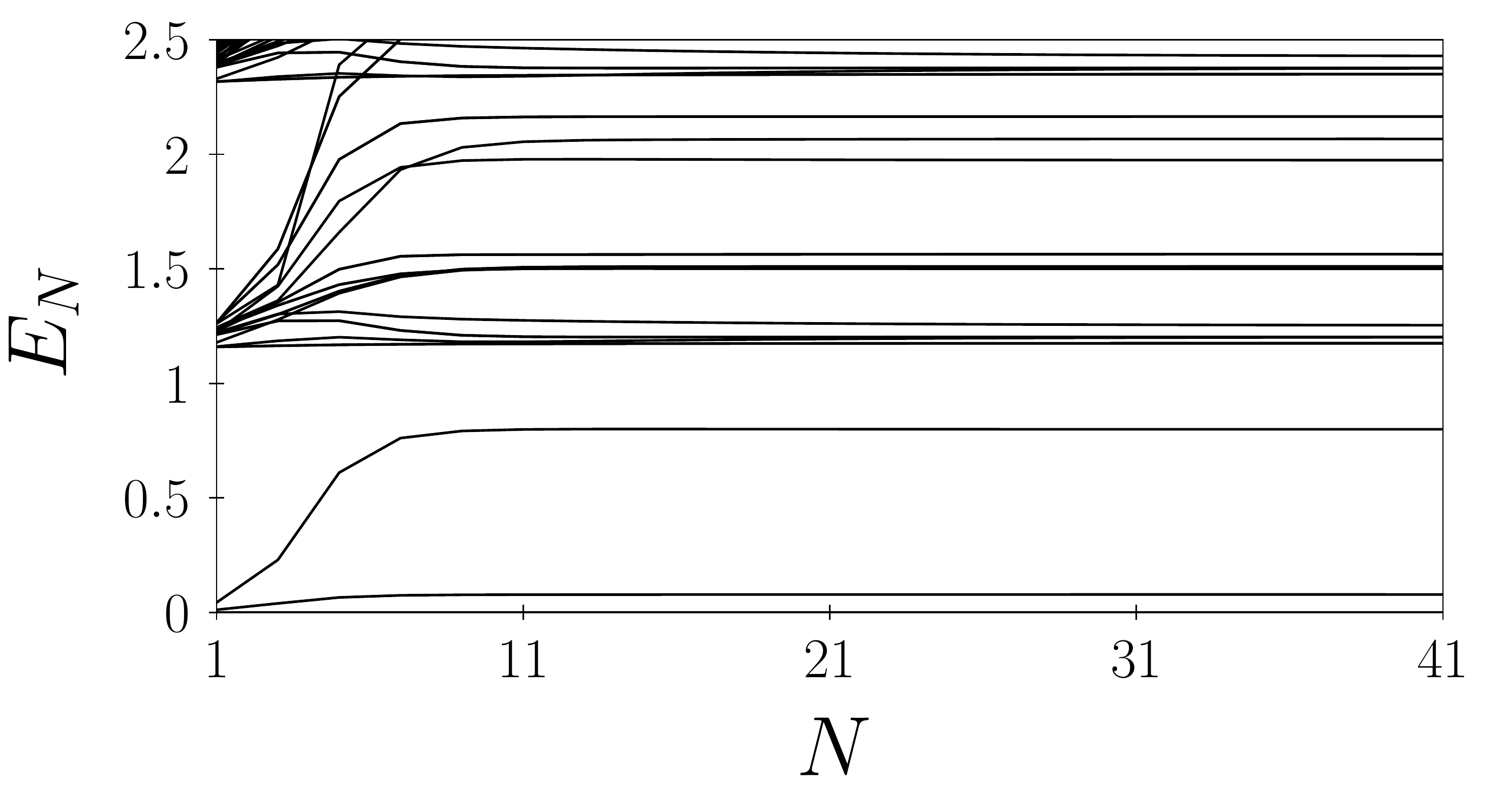}}
\caption{\label{fig:critical-spectrum}%
NRG critical spectrum of the Ising BFA model:
scaled energy $E(N)/D_N$ vs odd iteration $N$ for band exponent $r=0.6$, bath exponent $s=0.9$, and impurity parameters $U = 0.03D$, $\epsilon_d \simeq -0.025D$, $\Gamma=0.1 D$, and $g = 0.3D$. 
}
\end{figure}

Figure 1(a) of the main paper shows the evolution with NRG iteration number $N$ of the scaled energy $E_N$ of a single NRG many-body eigenstate of the Ising-anisotropic particle-hole-asymmetric BFA model, as calculated for different values of the bosonic coupling $g$ near its critical value $g_c$ on the boundary between the Kondo and localized (or Kondo-destroyed) phases. As $g$ approaches $g_c$, $E_N$ remains to ever higher iteration numbers close to a value specific to the spectrum of the Kondo-destruction quantum critical point. Figure \ref{fig:critical-spectrum} shows the critical spectrum $E_N$ vs $N$ for all states $E_N < 2.5 D_N$, where $D_N \simeq D\Lambda^{-N/2}$ is the characteristic energy scale of NRG iteration $N$. Since $|g/g_c - 1| < 10^{-13}$, any appreciable flow away from the QCP toward the Kondo or localized phase sets in at an iteration $N>41$.

\subsection{Additional CTQMC Results and Scaling Analysis}

Figures 1, 3, and 4 of the main paper present CTQMC results for the BFA model with SU(2) spin symmetry. This section presents the corresponding data for the Ising-anisotropic model, and provides additional discussion of the dynamical scaling in both the Ising and SU(2) cases.

\begin{figure}[h!]
\centering\mbox{\includegraphics[width=0.7\columnwidth]{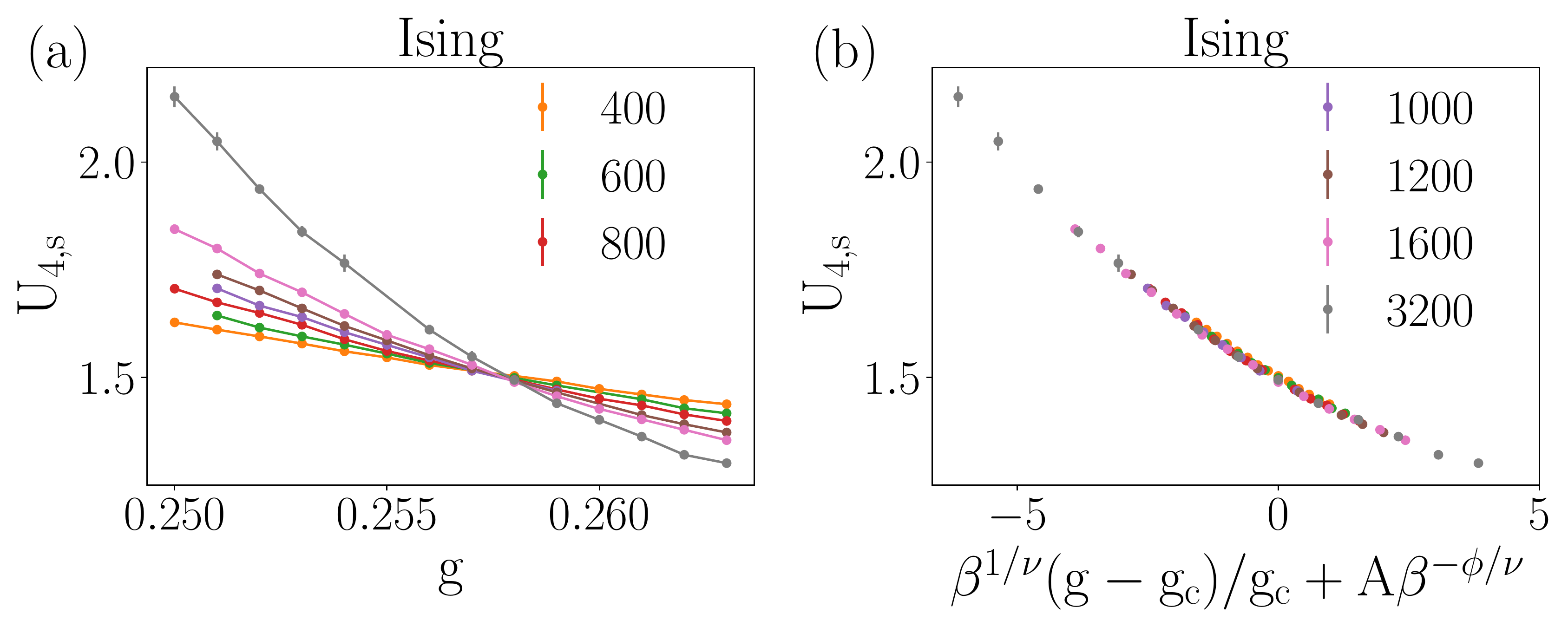}}
\caption{\label{fig:ising_u4}%
(a) Binder cumulant $U_{4,s}(\beta,g)$ vs bosonic coupling $g$ of the Ising data. (b) Scaling collapse of the same data optimized by $g_c = 0.258(1)$,
$\nu^{-1}=0.655(18)$, $A=0.0$, and $\phi=0.0$.}
\end{figure}

Figure \ref{fig:ising_u4}(a) plots the raw Binder cumulant $U_{4,s}$ in the Ising BFA model, showing curves for different temperatures crossing at a singe point. 
Figure \ref{fig:ising_u4}(b) demonstrates the scaling collapse of the raw data using optimized parameters specified in the legend.

\begin{figure}[h!]
\centering\mbox{\includegraphics[width=0.8\columnwidth]{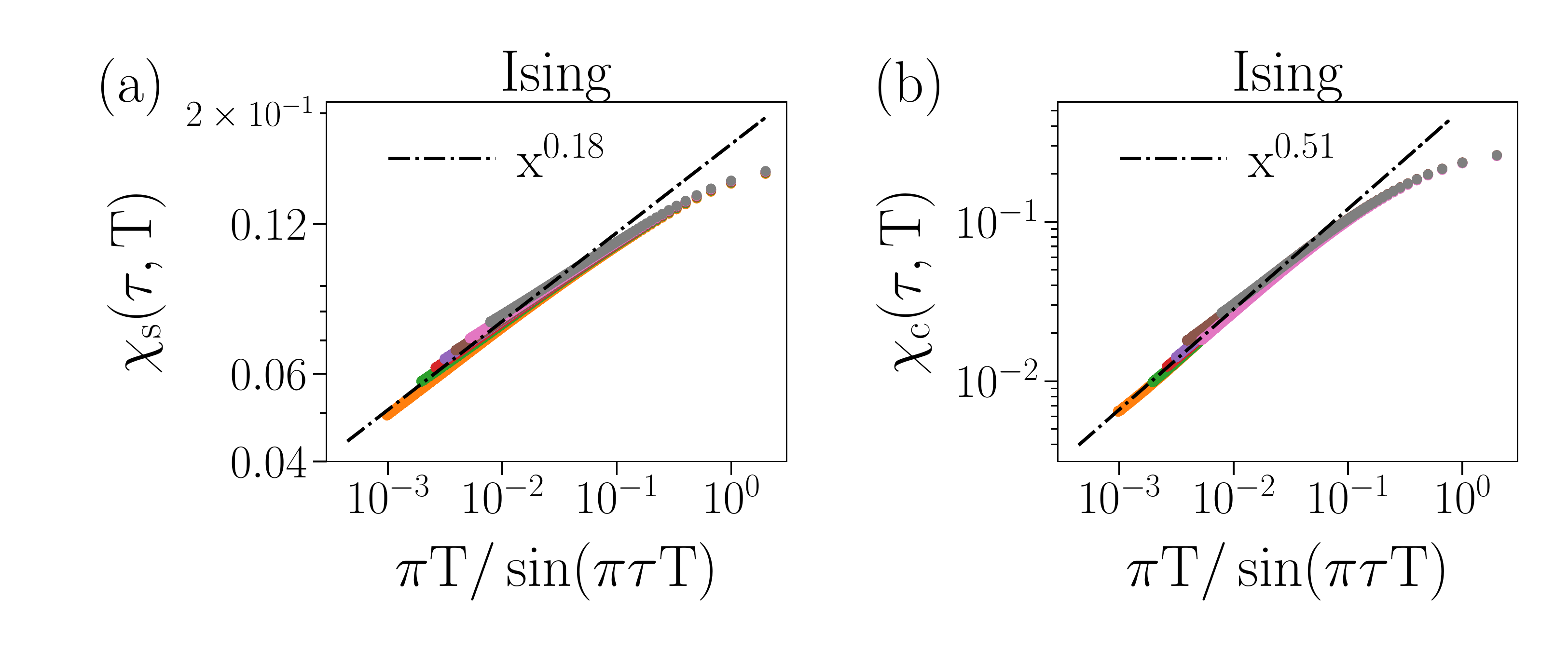}}    
\caption{\label{fig:dyn_scal_ising}%
(a) Spin and (b) charge dynamical local susceptibilities at the Kondo-destruction QCP in the Ising BFA model, showing scaling collapse of data for different inverse temperatures when plotted vs $\pi T / \sin(\pi\tau T)$.
Colors are as specified in the legends of Fig.\ \ref{fig:ising_u4}.
}
\end{figure}

Figure \ref{fig:dyn_scal_ising} shows the imaginary time dependence of the local spin and charge susceptibilities at the Kondo-destruction QCP. As also seen for the SU(2) BFA model in Figs.\ 3(b) and 4(b) of the main paper, the Ising-BFA data for different temperatures collapse onto the conformal scaling form
\begin{equation}
\chi_{\alpha}(\tau,T) = \Phi_\alpha \bigg[ \frac{\pi T}{\sin(\pi\tau T)} \bigg]
\quad \text{for } \alpha = s,\, c ,
\end{equation}
with long-imaginary-time behavior $\chi_\alpha \sim \tau^{-\eta_\alpha}$. The fitted dynamical exponents, $\eta_s = 0.1878(2)$ and $\eta_c = 0.5130(14)$, are respectively quite close to $1 - x_s = 0.20(1)$ and $\eta_c =1 - x_c = 0.53(2)$ computed over the same temperature range. Similarly, at the SU(2)-symmetric QCP, the fitted values $\eta_s = 0.2000(1)$ and $\eta_c = 0.5728(2)$ are close to $1 - x_s = 0.20(1)$ and $1 - x_c = 0.56(3)$. The fact that $\eta_c = 1 - x_c$ over a temperature window where there are significant corrections to the critical static $\chi_c$ indicates that both the leading and subleading terms in the critical $\chi_c(\tau,T)$ scale in terms of $\pi T/\sin(\pi\tau T)$.
This imaginary-time scaling implies $\omega/T$ scaling in the frequency domain.

\bibliography{pha-bfam}